\documentclass[useAMS,usenatbib]{mn2e}
\usepackage{graphicx,epsfig,amssymb,amsmath,layout,verbatim,rotating,calc,mathrsfs}
\usepackage{txfonts}
\def\mnras{MNRAS}
\def\aa{A\&A}
\def\aj{AJ}
\def\apj{ApJ}
\def\apjl{ApJL}
\def\cmda{Cel. Mech. Dyn. Astron.}
\title[The rate of secular evolution in elliptical galaxies with central
masses] {The rate of secular evolution in elliptical galaxies with
central masses}
\author[C. Kalapotharakos]{Constantinos Kalapotharakos\thanks{E-mail:ckalapot@phys.uoa.gr (CK)}\\
Academy of Athens, Research Center for Astronomy, 4 Soranou
Efesiou Str., GR-11527 Athens, Greece}

\date{Accepted ..........Received .............;in original form ..........}

\pagerange{\pageref{firstpage}--\pageref{lastpage}}

\pubyear{2007}

\begin{document}

\maketitle

\label{firstpage}

\begin{abstract}
We study a series of $N-$body simulations representing elliptical
galaxies with central masses. Starting from two different systems
with smooth centres, which have initially a triaxial configuration
and are in equilibrium, we insert to them central masses of
various values. Immediately after such an insertion a system
presents a high fraction of particles moving in chaotic orbits, a
fact causing a secular evolution towards a new equilibrium state.
The chaotic orbits responsible for the secular evolution are
identified. Their typical Lypaunov exponents are found to scale
with the central mass as a power law $L\propto m^s$ with $s$ close
to $1/2$. The requirements for an effective secular evolution
within a Hubble time are examined. These requirements are
quantified by introducing a quantity called \emph{effective
chaotic momentum} $\mathscr{L}$. This quantity is found to
correlate well with the rate of the systems' secular evolution. In
particular, we find that when $\mathscr{L}$ falls below a
threshold value (0.004 in our $N-$body units) a system does no
longer exhibit significant secular evolution.
\end{abstract}

\begin{keywords}
stellar dynamics -- methods: $N$-body simulations -- galaxies:
evolution -- methods: numerical -- galaxies: elliptical and
lenticular, cD -- galaxies: kinematics and dynamics
\end{keywords}
%

\section{Introduction}
Schwarzschild's (1979) pioneering method confirmed the existence
of self-consistent models of elliptical galaxies with exclusively
regular orbits (Schwarzschild 1979, 1982; Richstone 1980, 1982,
1984; Richstone \& Tremaine 1984; Levison \& Richstone 1987).
This, together with the study of the perfect ellipsoid model (de
Zeeuw \& Lyndel-Bell 1985; Statler 1987), which is fully
integrable, contributed to the formation of a prevailing aspect
during the '80s that the elliptical galaxies in equilibrium mainly
consist of stars moving in regular orbits, while stars in chaotic
orbits are either few or inexistent. For this reason galactic
dynamical studies were based for many years on a systematic
exploration of the stable families of periodic orbits in static
galactic potentials.

\begin{figure*}
\label{figphsporb}
\centerline{\includegraphics[width=14cm]{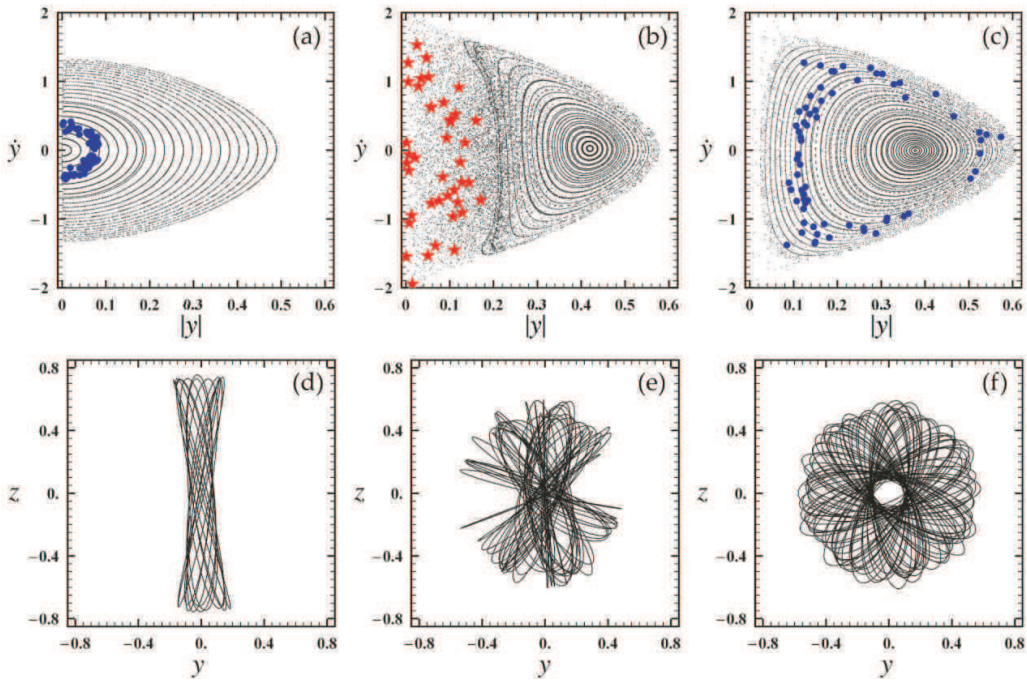}}
\caption{\textbf{a)} The phase portrait (projections on ($|y|,
\dot{y}$) of the 4D Poincar\'{e} surface of sections ($z=0,
\dot{z}>0$)) at an energy level $\simeq 0.74\times$ (the central
potential well) in a typical case of our $N-$body simulations (see
below), before the insertion of the CM. We see the invariant curves
of box orbits. \textbf{b)} The same phase portrait at $t=90T_{hmct}$
after the insertion of the CM. The invariant curves corresponding to
box orbits have been destroyed. At their place we have a chaotic
domain and an island of regular short axis tube orbits. As the
system evolves the chaotic domain is reduced gradually and the
island of SAT orbits grows. \textbf{c)} At $t=300T_{hmct}$ after the
insertion of the CM, the system has reached its final equilibrium
state. The SAT island prevails. In each of these panels blue dots
correspond to regular motion while red stars correspond to chaotic
motion of one of the $N-$body particles. \textbf{d)} Before the
insertion of the CM the particle moves in a box orbit. \textbf{e)}
For $t=0-180T_{hmct}$ after the insertion of the CM the particle
wanders in the chaotic domain. \textbf{f)} After $t\simeq
180T_{hmct}$ the particle is trapped by SAT tori and its orbit is
converted to regular, of the SAT type.}
\end{figure*}

This point of view was challenged in the early '90s when
Schwarzschild (1993) found that chaos plays a significant role in
the structure of systems with cuspy density profiles
$(\rho_{halo}\propto r^{-2})$. A number of observations also
indicated that high values of \textbf{C}entral \textbf{M}asses
(CM) exist in galaxies (Crane et al. 1993; Ferrarese et al. 1994;
Lauer et al. 1995; Kormendy \& Richstone 1995; Gebhardt et al.
1996; Faber et al. 1997; Kormendy et al. 1997, 1998; van der Marel
et al. 1997; van der Marel \& van den Bosch 1998; Magorrian et al.
1998; Cretton \& van den Bosch 1999; Gebhardt et al. 2000).
Gerhard \& Binney (1985) had shown how box orbits are converted to
chaotic orbits in systems with strong CM. Later, many authors
studied in detail this phenomenon and its consequences for
elliptical galaxies (Merritt \& Fridman 1996; Merritt \& Valuri
1996, 1999; Fridman \& Merritt 1997, Valluri \& Merritt 1998;
Merritt \& Quinlan 1998; Siopis 1999, Siopis \& Kandrup 2000;
Holley­Bockelmann et al. 2001, 2002; Poon \& Merritt 2001, 2002,
2004; Kandrup \& Sideris 2002; Kandrup \& Siopis 2003;
Kalapotharakos, Voglis \& Contopoulos 2004; Kalapotharakos \&
Voglis 2005, see Merritt 1999, 2006 and Efthymiopoulos, Voglis \&
Kalapotharakos 2007 for a review). In particular, using the
Schwarzschild's method Merritt \& Fridman (1996) found
self-consistent solutions for triaxial configurations in the case
of a weakly cuspy profile $\rho\propto r^{-1}$. In these solutions
the fraction of particles in chaotic orbits raised up to
$\simeq46\%$. On the other hand, in the case of strongly cuspy
profiles $\rho\propto r^{-2}$ the optimal solutions found had
$\simeq60\%$ chaotic orbits. However, the solutions in the latter
case were not fully stationary.

It should be noted that, while a solution of the Schwarzschild
method should ideally represent a stationary galaxy, in practice
an $N-$body realization of such a solution will show some secular
evolution due to a number of reasons, namely \textbf{a)} the
solution's response density may not match precisely the imposed
density \textbf{b)} some of the chaotic orbits of the solution may
not be `fully mixed' i.e. they may have not reached a full
covering of their invariant measure in the phase space within the
method's fixed integration time, and \textbf{c)} by its nature
Schwarzschild's method cannot secure the stability of even a
perfect solution. Thus, questions related to the \emph{stability}
or \emph{secular evolution} of a self-consistent model of a galaxy
are better answered by the \emph{$N-$body} method, which is free
of the limitations of Schwarzschild's method.

Many works by the $N-$body method have addressed the questions of
the orbital analysis and of the relation between the
self-consistency and the orbital content of either stationary or
evolving systems (Holley-Bockelmann et al. 2001, 2002;
Contopoulos, Voglis \& Kalapotharakos 2002; Voglis, Kalapotharakos
\& Stavropoulos 2002; Kalapotharakos et al. 2004; Kalapotharakos
\& Voglis 2005; Muzzio, Carpintero \& Wachlin 2005; Jesseit, Naab
\& Burkert 2005; Muzzio 2006; Voglis, Stavropoulos \&
Kalapotharakos 2006). In these works it has been repeatedly
demonstrated that the insertion of a CM in a triaxial system at
equilibrium leads to significant secular evolution, resulting in a
new equilibrium state which is, in most cases oblate (Merritt \&
Quinlan 1998; Holley-Bockelmann et al. 2001, 2002; Kalapotharakos
et al. 2004; Kalapotharakos \& Voglis 2005).

Kalapotharakos et al. (2004), Kalapotharakos \& Voglis (2005)
extended the research of Voglis et al. (2002) for a series of
models with CMs. These works unravelled very clearly the
transformation of the phase space during the secular evolution that
leads the systems to a new equilibrium state. A summary of our
previous results, needed in the sequel, can be made with the help of
Fig.~1. Fig.~1a shows the phase space structure in an example of our
simulations before the insertion of the CM. The phase portrait is
almost entirely filled by invariant curves corresponding to box
orbits. The big blue dots correspond to the Poincar\'{e} consequents
of a specific particle's orbit (called hereafter particle A). Note,
that these dots are provided by the self-consistent $N-$body run.
After the insertion of the CM practically all the tori of box orbits
are destroyed and their place is occupied by a chaotic domain and by
smaller islands corresponding to the well known `boxlets' (e.g.
Merritt 1999; these orbits were called \textbf{H}igh \textbf{O}rder
\textbf{R}esonant \textbf{T}ube (HORT) in previous works of ours).
Now, as the chaotic orbits diffuse in the phase space they cause a
secular change of the system's form, which becomes more spherical
and less prolate. This change, on its turn, affects the phase space
favouring \textbf{S}hort \textbf{A}xis \textbf{T}ube (hereafter SAT)
orbits. Fig.~1b shows the phase space structure (black points) at
$t=90$ half mass crossing times (hereafter $T_{hmct}$). We observe a
large chaotic domain (for $|y|\lesssim 0.2$) and a big island of
stability (for $|y|\gtrsim 0.2$) corresponding to SAT orbits. Note,
that in the present paper we always consider the half mass radius
$R_h$ being equal to unity ($R_h=1$). The red stars correspond to
the Poincar\'{e} consequents of the orbit of particle A in the time
interval $t=0-180T_{hmct}$. The main effect is the following: As the
island of SAT orbits grows, the majority of particles, initially in
chaotic orbits, are gradually trapped in the domain of SAT tori and
the orbits are converted to regular, of the SAT type. Fig.~1c shows
this conversion for particle A. The second row of Fig.~1 shows also
the gradual conversion of the same orbit from box to chaotic and
then to SAT, as it appears in ordinary space. The phase space
structure in Fig.~1c corresponds to $t=300T_{hmct}$ (a Hubble time)
and the big blue dots correspond to the consequents of the orbit of
particle A within the time interval $t=180-350T_{hmct}$. At
$t=300T_{hmct}$ the system has already reached its final equilibrium
state. At this snapshot the number of particles in chaotic orbits is
considerably lower than at $t=0$.

In the systems studied by Kalapotharakos et al. (2004) the
destruction of box tori after the insertion of the CM results in
the fraction of mass in chaotic orbits rising initially up to the
level of $50\%-80\%$ (depending on the initial maximum ellipticity
of the system). However, during the secular evolution of these
systems the fraction of chaotic orbits decreases significantly and
at the new equilibrium state this fraction ranges between
$\simeq10\%$ and $\simeq25\%$. The rate of secular evolution
depends on the mass value of the CM as well as on the
configuration of the original system before the insertion. In
particular, different mass values lead not only to different
fractions of particles in chaotic orbits, but also to different
distributions of the particles' Lyapunov exponents.

In the present paper our goals are:

\textbf{a)} to study the relation between the mass value of the CM,
on the one hand, and the corresponding values of the Lyapunov
characteristic exponents, on the other hand, of those chaotic orbits
which are produced due to the insertion of the CM and are
responsible for the secular evolution. We can immediately state the
result of this investigation, which constitutes one principal result
of the present paper: we find that this relation is a power law with
an exponent close to $1/2$. Work on the theoretical justification of
such a scaling law is in progress, but a preliminary discussion is
given in the final section of the paper.

\textbf{b)} to determine the specific requirements so that a
system exhibits significant secular evolution within a Hubble
time. Surprisingly, it was found that, independently of the
details of a system, the onset of significant secular evolution
takes place when a quantity termed \emph{effective chaotic
momentum} surpasses a threshold value (equal to 0.004 in the
present paper's $N-$body units). This threshold is global, i.e.
the same for all the studied systems. This indicates that the
effective secular momentum is a quantity possibly related to a
more fundamental statistical mechanical description of the systems
under study.

Section 2 briefly describes the models. Section 3 discusses the
relation between the value of the CM and the characteristic values
of the Lyapunov exponents of its associated chaotic orbits.
Section 4 contains the definition of the effective chaotic
momentum, that measures the ability of a system to undergo secular
evolution within a Hubble time. Finally, section 5 summarizes the
main conclusions of the present study.

\begin{table}
\caption{The fractions of the total number of particles in chaotic
orbits in the various systems for $t=0$ (the moment when the CM is
inserted).} \centering \label{tabt000}
\begin{tabular}{@{}lccccc@{}}\hline 
  models  & $m=0$ & $m=0.0005$ & $m=0.001$ & $m=0.005$ & $m=0.01$  \\ \hline
  Q & 32\% & 79\% & 82\% & 81\% & 80\% \\ 
  C & 23\% & 48\% & 51\% & 53\% & 50\% \\ \hline
\end{tabular}
\end{table}

\begin{table}
\caption{The fractions of the total number of particles in chaotic
orbits in the various systems for $t=150T_{hmct}$ (half a Hubble
time).} \centering \label{tabt150}
\begin{tabular}{@{}lccccc@{}}\hline 
  models  & $m=0$ & $m=0.0005$ & $m=0.001$ & $m=0.005$ & $m=0.01$  \\ \hline
  Q & 32\% & 78\% & 80\% & 73\% & 56\% \\ 
  C & 23\% & 47\% & 48\% & 40\% & 30\% \\ \hline
\end{tabular}
\end{table}

\begin{table}
\caption{The fractions of the total number of particles in chaotic
orbits in the various systems at the final equilibrium state.}
\centering \label{tabteq}
\begin{tabular}{@{}lccc@{}}\hline 
  models  & $m=0$ & $m=0.005$ & $m=0.01$  \\ \hline
  Q & 32\% & 25\%~(at $t=300$) & 22\%~(at $t=2200$) \\ 
  C & 23\% & 19\%~(at $t=2200$) & 12\%~(at $t=2200$) \\ \hline
\end{tabular}
\end{table}

\begin{figure}
\label{figq100chaost}
\centerline{\includegraphics[width=8.5cm]{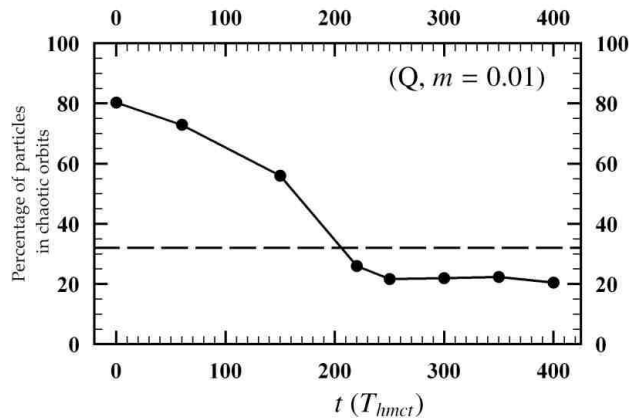}} \caption{The
evolution of the percentage of chaotic orbits in the system
(Q,~$m=0.01$). Initially, we have a high percentage ($\simeq 80\%$)
of chaotic orbits that decreases with time. When the system reaches
the final oblate equilibrium state the percentage of chaotic orbits
remains stabilized at a level $\simeq 20\%$. The horizontal dashed
line denotes the percentage of chaotic orbits in the Q system
(before the insertion of the CM).}
\end{figure}

\begin{figure*}
\label{figqlcndist}
\centerline{\includegraphics[width=14cm]{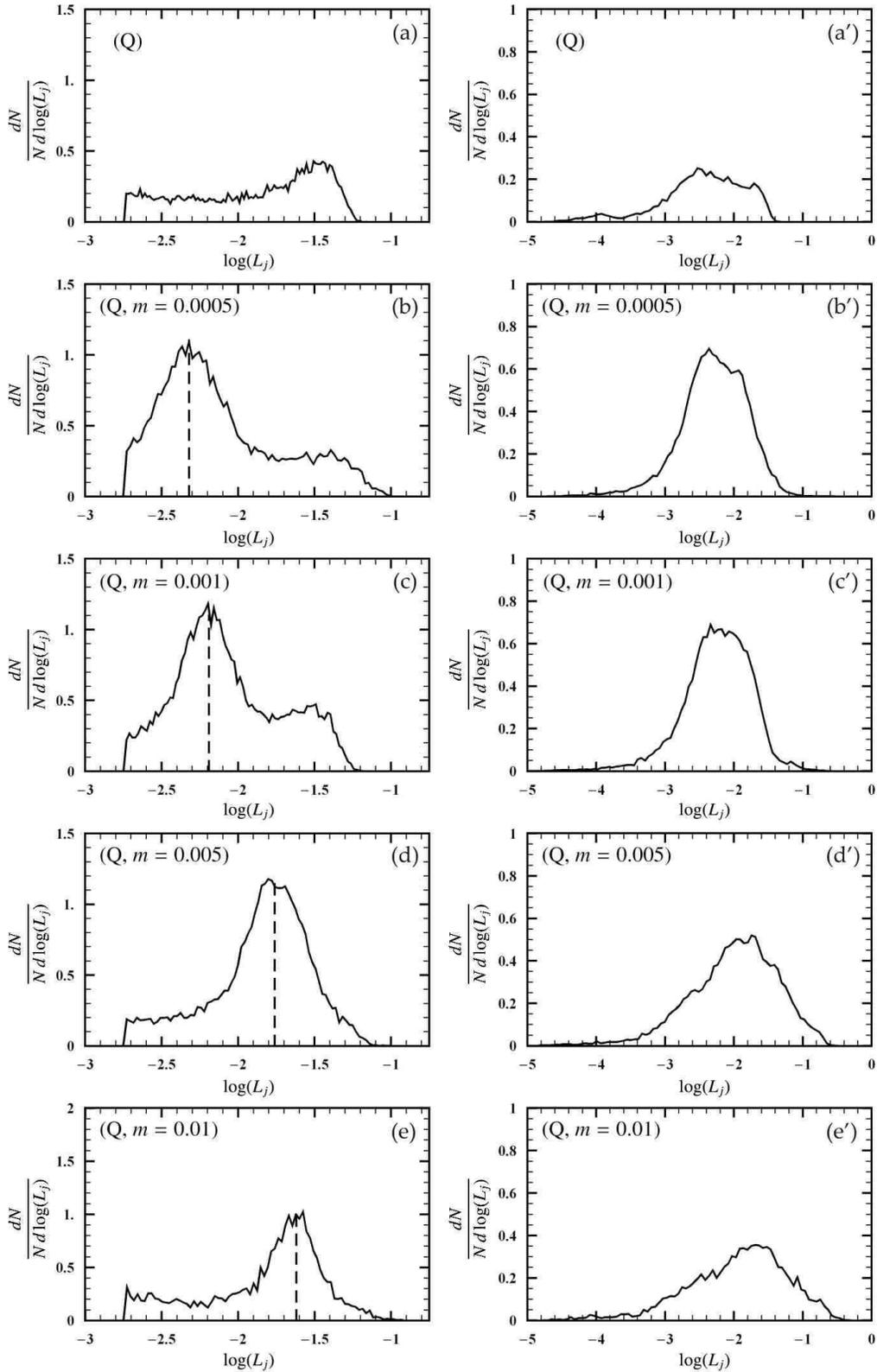}}
\caption{\textbf{Left column:} The distributions of $\log(L_j)$ for
the chaotic orbits of the Q family systems at the snapshot
$t=150T_{hmct}$, with mass parameters $m$ as indicated in the
panels. A pronounced maximum appears for the systems with $m\neq 0$,
which is shifted towards higher values of $\log(L_j)$ as $m$
increases. Note, that the area below each curve denotes the
corresponding function of chaotic orbits. \textbf{Right column:}
Same as in the left column for the distributions of
$\log(L_{cuj})$.}
\end{figure*}

\begin{figure*}
\label{figclcndist}
\centerline{\includegraphics[width=14cm]{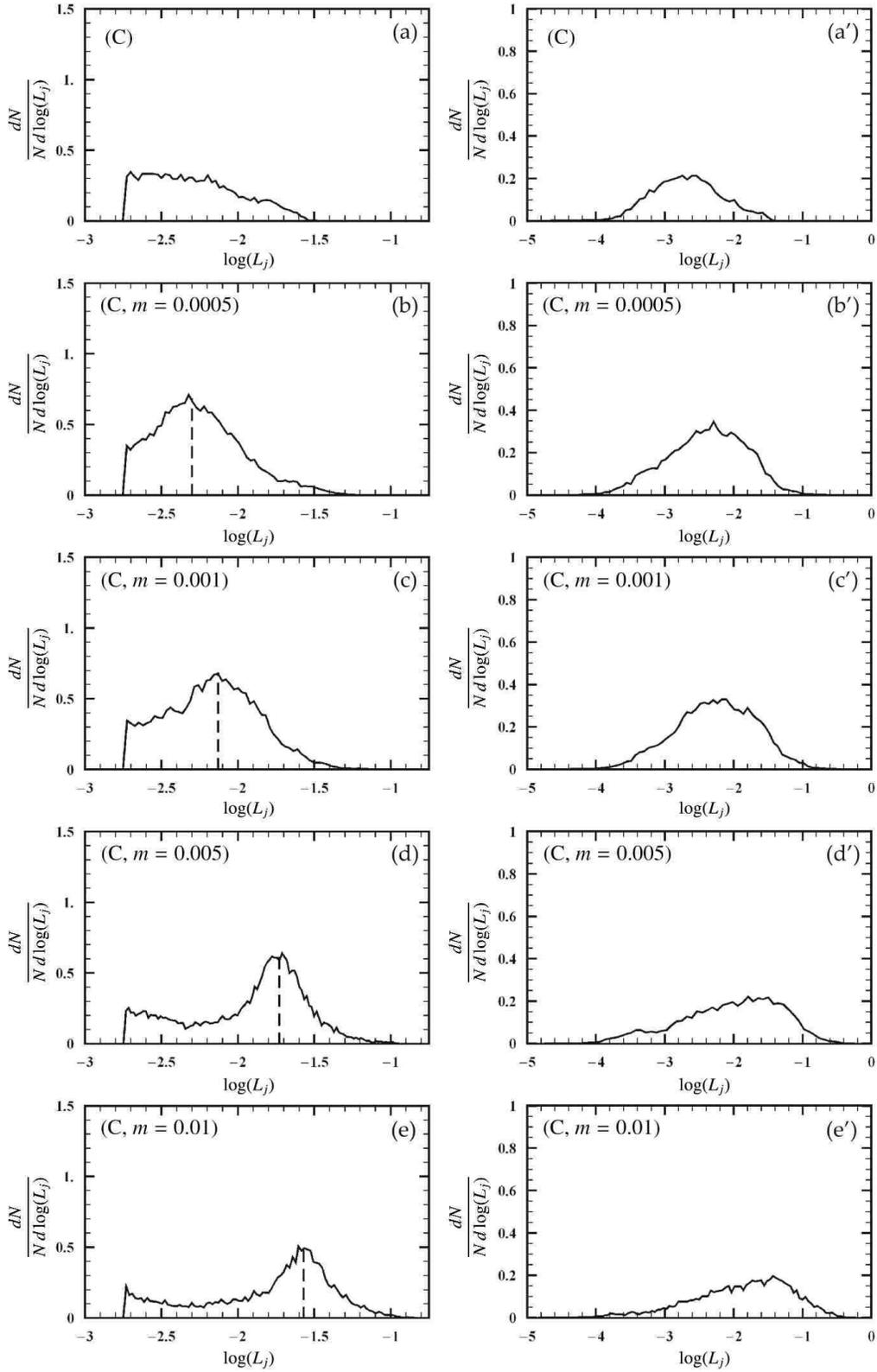}} \caption{Same
as in Fig.~3 but for the C family systems.}
\end{figure*}

\begin{figure*}
\label{figpowlaw}
\centerline{\includegraphics[width=17cm]{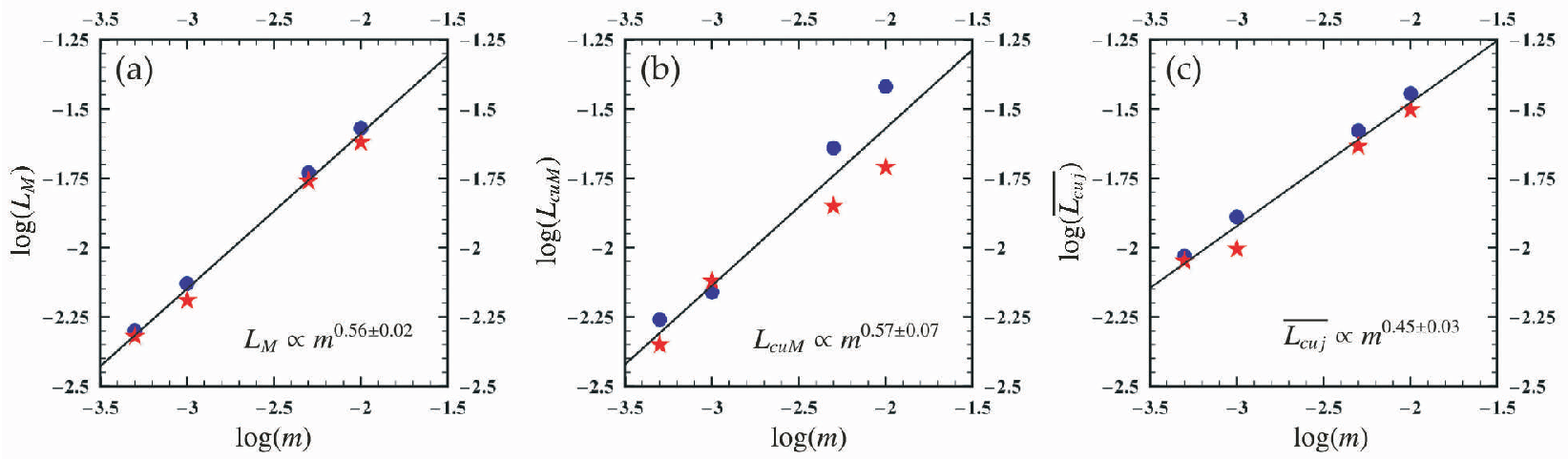}} \caption{The
values of \textbf{a)} $\log(L_M)$, \textbf{b)} $\log(L_{cuM})$ and
\textbf{c)} $\log(\overline{L_{cuj}})$ against $\log(m)$. Red stars
and blue dots correspond to the Q and C family systems respectively.
All three Lyapunov numbers
$-~\log(L_M)$,~$\log(L_{cuM})$,~$\log(\overline{L_{cuj}})~-$ are
well fitted by power laws $\propto m^s$ with exponents s near the
value 0.5.}
\end{figure*}

\section{Description of the models}

Our study refers to the insertion of a CM in a variety of
equilibrium $N-$body systems. The latter are described in detail
in Kalapotharakos et al. (2004). Here we describe briefly the most
important features of these systems.

Two families of models are examined in the present paper. They are
produced from two different original systems representing smooth
centre elliptical galaxies in equilibrium. The two initial systems
are both triaxial, but nearby prolate. The first system (called Q
system) has a maximum ellipticity $E_{max}\simeq 7$ and
triaxiality $Ô\simeq 0.9$ while the second system (called C
system) has a maximum ellipticity $E_{max}\simeq 3.5$ and
triaxiality $Ô\simeq 0.8$. Note that the short, intermediate and
long axes of each system in the present study is aligned along the
$x,y,z$ axes, respectively. Voglis et al. (2002) calculated the
fractions of particles moving in chaotic orbits in these systems.
They found $\simeq 32\%$ and $\simeq 23\%$ for the Q and C system,
respectively. These fractions, as well as the identities of the
particles moving in chaotic orbits, remain almost unchanged in
time, before the insertion of the CM.

We insert, in these systems, CMs with a variety of mass values and
follow numerically their evolution by the $N-$body code (Allen,
Palmer \& Papaloizou 1990). We then calculate the fractions of the
particles moving in chaotic orbits at various snapshots of this
evolution. The CM is assumed to have the following density profile
(Allen et al. 1990):
\begin{equation}
\label{eqrhocmc} \rho_{cm}=\dfrac{GM_{cm}a^2}{2\pi r(r^2+a^2)^2}
\end{equation}
where $a=0.005 m R_g$, $m=\dfrac{M_{cm}}{M_g}$ is the relative
value of the CM with respect to the total galaxy mass $M_g$, and
$R_g$ is the radius of galaxy. The profile \eqref{eqrhocmc} yields
a cuspy profile $r^{-1}$ for $r\ll a$. We examined four different
values of $m$, namely $(0.0005, 0.001, 0.005, 0.01)$.

The fraction of mass in chaotic motion, corresponding to the
integration of orbits in a `frozen' potential of each system at
the snapshot $t=0$ (the moment when the CM is inserted), is
approximately $80\%$ in all the Q models (for various masses $m$),
and approximately $50\%$ in all the C models. These percentages
turn to be rather irrelevant to the values of $m$ considered
(Table \ref{tabt000}).

In subsequent (isochronous) snapshots, these fractions are seen to
vary for different values of $m$. Table \ref{tabt150} shows the
fractions of particles detected in chaotic orbits at the time
$t=150T_{hmct}$ (half a Hubble time) after the insertion of the
CM. The main remark with respect to this table is that, the
systems with higher values of $m$ (0.005, 0.01) exhibit
\emph{smaller} fractions of chaotic orbits than the systems with
lower values of $m$ (0.0005, 0.001) which retain a fraction of
chaotic orbits almost equal to the fractions at the snapshot $t=0$
(compare Tables \ref{tabt000} and \ref{tabt150}).

Now, Table \ref{tabteq} shows the fractions of the chaotic orbits
for the systems with high values of $m$ (0.005, 0.01), when the
systems have reached a new equilibrium state. The main observation
here is that in all cases the fraction of chaotic orbits at the
new equilibrium state is far smaller than the fraction of the
chaotic orbits in the same system at $t=0$ (compare Tables
\ref{tabt000} and \ref{tabteq}). In fact, the final fraction is
even smaller than the fraction of chaotic orbits in the same
systems before the insertion of the CM. In Fig.~2 we see, for
example, the evolution of the fraction of chaotic orbits for the
system (Q,~$m=0.01$). The gradual decrease of this fraction is
shown clearly up to the moment when the system reaches the new
oblate equilibrium state ($t\simeq 220$). In the case of systems
with low values of $m$ the corresponding fractions were not
calculated because these systems need much longer intervals than a
Hubble time to reach an equilibrium. For example, after a period
of 20 Hubble times the $m=0.001$ system of the Q family appears to
be still far from equilibrium (see fig.~4 of Kalapotharakos et al.
2004). The times (in $T_{hmct} units$) of final states of all the
systems that have reached a final equilibrium state is given in
Table 3, for comparison.

\begin{figure*}
\label{figq100t150}
\centerline{\includegraphics[width=17cm]{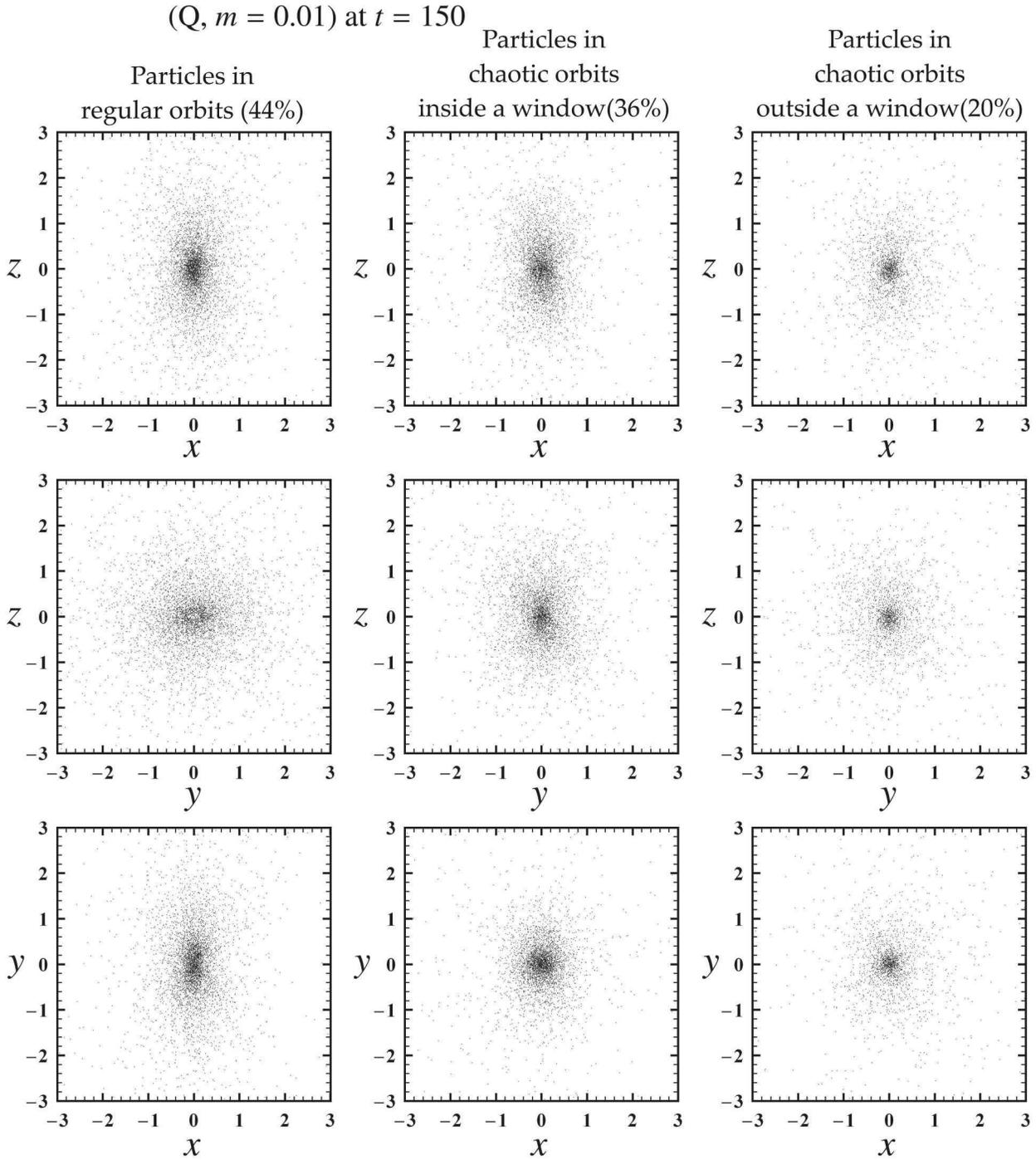}} \caption{Each
column shows the projections of different mass components of the
(Q,~$m=0.01$) system on the three principal planes ($x-z,~y-z,~x-y$)
at $t=150T_{hmct}$. \textbf{Left column:} Projection of the
particles in regular orbits. These particles constitute
approximately $44\%$ of the total mass and they form a strongly
flattened (oblate) configuration. The orbits are mainly of SAT and
boxlet (called `HORT' in Kalapotharakos \& Voglis 2005, see text)
type. \textbf{Middle column:} Projection of the particles in chaotic
orbits with $L_j$s within a window located around the maximum
$\log(L_M)$ of the distribution of Fig.~3e. These particles
constitute approximately $36\%$ of the total mass. They form a
triaxial configuration and they were moving in box orbits before the
insertion of the CM. After the insertion, they are responsible,
through their chaotic diffusion, for the secular evolution exhibited
by the system. \textbf{Right column:} Projection of the particles in
chaotic orbits with $L_j$s outside the window around the $\log(L_M)$
maximum. This component constitutes approximately $20\%$ of the
total mass and it forms an isotropic (spherical) configuration. The
spherical distribution is due to the nearly complete mixing of the
particles in the phase space due to chaotic diffusion. Such
particles no longer contribute to the systems's overall secular
evolution.}
\end{figure*}

\section{CM and Lyapunov exponents}
In order to distinguish the orbits into regular and chaotic we use
two indices: \textbf{a)} The Specific Finite Lyapunov
Characteristic Number or $L_j$ and \textbf{b)} the Alignment Index
(AI) or Smaller ALignment Index (SALI) (for more details see
Voglis, Contopoulos \& Efthymiopoulos 1998; Voglis, Contopoulos \&
Efthymiopoulos 1999; Skokos 2001; Voglis et al. 2002). The $L_j$
has as unit the inverse of the radial period of each orbit
$T_{rj}$ and provides a good answer regarding whether an orbit can
be characterized as regular or chaotic (up to a certain level). It
can also provide a comparison of the relative degree of chaos of
two orbits provided that their $L_j$s have already reached a
stabilized value within a given integration time. However, since
the $L_j$ is measured by a different unit for each orbit, it
cannot be used for a comparison of the orbits as regards the
efficiency of their chaotic diffusion in phase space. For this
reason, the $L_j$s are converted to quantities with a common unit
for all the particles. The Common Unit Finite Time Lyapunov
Characteristic Number or simply $L_{cuj}$ is defined by
\begin{equation}
\label{eqlcuj} L_{cuj}=L_j\dfrac{T_{hmr}}{T_{rj}}
\end{equation}
in units $1/T_{hmr}$, where $T_{hmr}$ is the period of radial
oscillations of the orbits with energy equal to the value of the
potential at a radius equal to the half-mass radius $R_h$. One
finds $T_{hmr}\simeq 3 T_{hmct}$. This Lyapunov exponent $L_{cuj}$
compares the chaotic orbits with respect to their ability to
undergo significant chaotic diffusion within a Hubble time. In
summary, the value of $L_j$ gives us a measure of the overall
complexity and chaoticity of the phase space inside which the
orbits lie, while the value of $L_{cuj}$ yields the exponential
rate of deviation, measured in absolute time units for each
individual orbit.

\begin{figure*}
\label{figq100t300}
\centerline{\includegraphics[width=10cm]{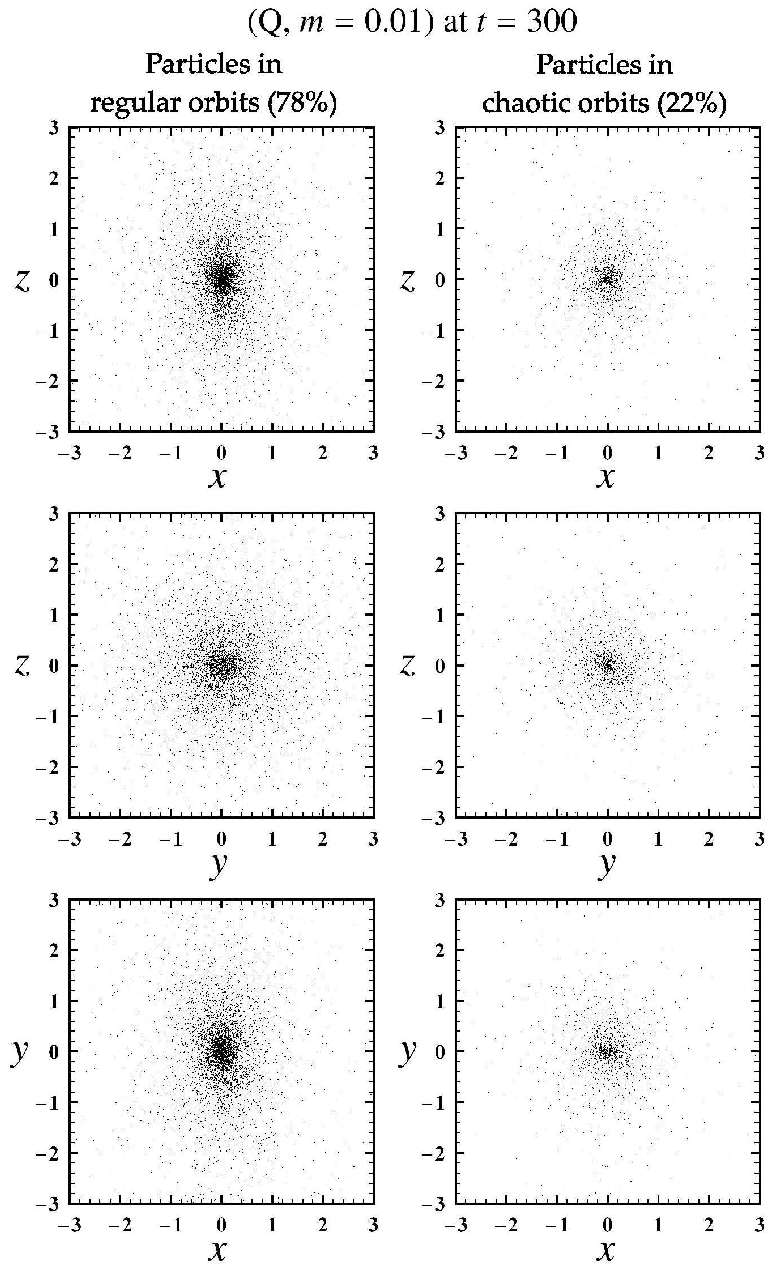}}
\caption{\textbf{Left column:} The projections of the particles in
regular orbits of (Q,~$m=0.01$) system on the three principal planes
at $t=300T_{hmct}$. These particles constitute $78\%$ of the total
mass and form an oblate configuration. \textbf{Right column:} The
same projection of the particles moving in chaotic orbits. These
particles constitute $22\%$ of the total mass and form an almost
spherical configuration. The chaotic diffusion does no longer have
any consequences on the form of the system.}
\end{figure*}

Now, Fig.~3 shows the distributions of the values of $\log(L_j)$
axis (left column) or the values of $\log(L_{cuj})$ (right column)
at the snapshot $t=150T_{hmct}$ for the particles in chaotic
orbits of all the systems of the Q family. The distributions are
normalized with respect to the systems' fraction of chaotic
orbits, i.e., the area below each curve denotes the fraction of
chaotic orbits. The following are observed:

\textbf{a)} All the distributions of the values of $\log(L_j)$ for
$m>0$ (Fig.~3 - left column) exhibit a global maximum at values
$\log(L_j)=\log(L_M)$. For $m=0$ (Fig.~3a) there appears no
conspicuous maximum. On the other hand, for $m>0$ there appears a
conspicuous maximum which is, furthermore, shifted to the right as
$m$ increases. By examining many orbits with $L_j$s in an interval
around these maxima we found that such orbits correspond to
particles that were initially moving in box orbits (before the
insertion of the CM).

\textbf{b)} The distributions of the values of $\log(L_{cuj})$
axis (Fig.~3 - right column) are quite different from the
corresponding distributions of $\log(L_j)$. Nevertheless, these
distributions also exhibit global maxima, at values
$\log(L_{cuM})$ not very different from $\log(L_M)$.

Similar remarks can be made for the systems of the C family
(Fig.~4). It is clear from Figs.~3 and 4 that the values
$\log(L_M)$ and $\log(L_{cuM})$ (for $m\neq 0$) depend on $m$. The
mean value of $L_{cuj}$, denoted by $\overline{L_{cuj}}$, also
depends on $m$. The main result now, is that all the three
dependencies are power laws functions of $m$ (Fig.~5). In Fig.~5,
the red stars refer to the systems of the Q family and the blue
dots to the systems of the C family. The power laws of $L_M$
(Fig.~5a), $L_{cuM}$ (Fig.~5b) and $\overline{L_{cuj}}$ (Fig.~5c)
are
\begin{subequations}
\begin{align}\label{eqpowlaw}
L_M\propto m^{s_1}\\
L_{cuM}\propto m^{s_2}\\
\overline{L_{cuj}}\propto m^{s_3}
\end{align}
\end{subequations}
where $s_1=0.56\pm 0.02,~s_2=0.57\pm 0.07$ and $s_3=0.45\pm 0.03$.
The quantities $L_M$, $L_{cuM}$ and $\overline{L_{cuj}}$ are
different measures of the mean level of the Lyapunov exponents of
the orbits, which have become chaotic, precisely by the insertion
of the CM. In conclusion, the \emph{Lyapunov characteristic
exponents of the chaotic orbits scale approximately as} $m^{1/2}$.
This property appears here as a numerical result, but it probably
has some theoretical explanation related to the scattering of the
orbits by the CM near the centre (see discussion).

\section{Effective chaotic momentum}
In Fig.~5 we see that the systems with $m<0.0050$, have an
$\overline{L_{cuj}}$ close to the value $10^{-2}$. These systems
do not exhibit appreciable evolution within a Hubble time. On the
other hand, the systems with $m\geq 0.0050$, develop a well
detectable secular evolution within a Hubble time. In that case
the $\overline{L_{cuj}}$ values are close to $10^{-1.5}$. In a
first approximation, we can say that this difference between the
mean Lyapunov exponents of the systems with low or high central
mass value $m$ can explain the faster evolution rate in the latter
case, since the diffusion of a chaotic orbit is effective within a
Hubble time ($\sim 100 T_{hmr}$) only when its $L_{cuj}$ value is
$L_{cuj}\gtrsim 10^{-2}$.

However, a more careful study revealed two basic requirements
which must be fulfilled in order for a system to undergo
significant secular evolution within a Hubble time:

\begin{enumerate}
    \item the \emph{fraction} of mass in \emph{chaotic motion} with
high $L_{cuj}$ values ($L_{cuj}>10^{-2}$) must not be very small
and
    \item this mass must have an \emph{anisotropic} initial distribution
    in ordinary space. In fact, whenever the mass appears isotropically distributed
    (i.e. close to a spherical distribution) this is an indication
    that the chaotic mixing has already taken place in the phase
    space, i.e. the particles in chaotic orbits have already acquired
    an almost uniform distribution in the phase space. We should
    stress that a spherical distribution of the particles in
    \emph{chaotic} motion is not incompatible with the appearance of chaos itself, since
    the particles in \emph{regular} orbits provide a significant
    perturbation of the overall galactic potential from a
    spherical potential.
\end{enumerate}

\begin{figure}
\label{figq100lcndistt}
\centerline{\includegraphics[width=8.5cm]{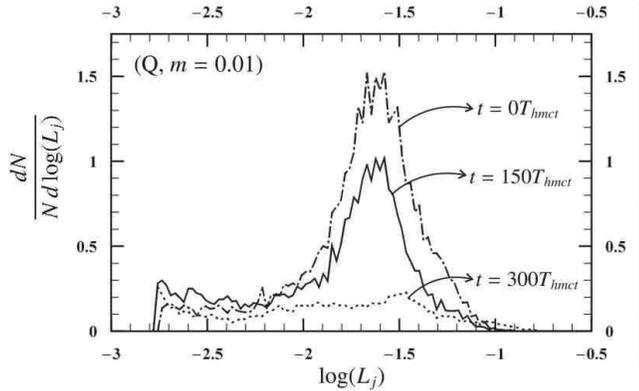}} \caption{The
distribution of $\log(L_j)$  for the particles in chaotic orbits of
the system (Q,~$m=0.01$) at three different times \textbf{a)} $t=0$,
\textbf{b)} $t=150T_{hmct}$ \textbf{c)} $t=300T_{hmct}$. The mass
converted from chaotic to regular orbits during the secular
evolution comes mainly from the area around the global maximum.}
\end{figure}

\begin{figure}
\label{figinerttens}
\centerline{\includegraphics[width=8.5cm]{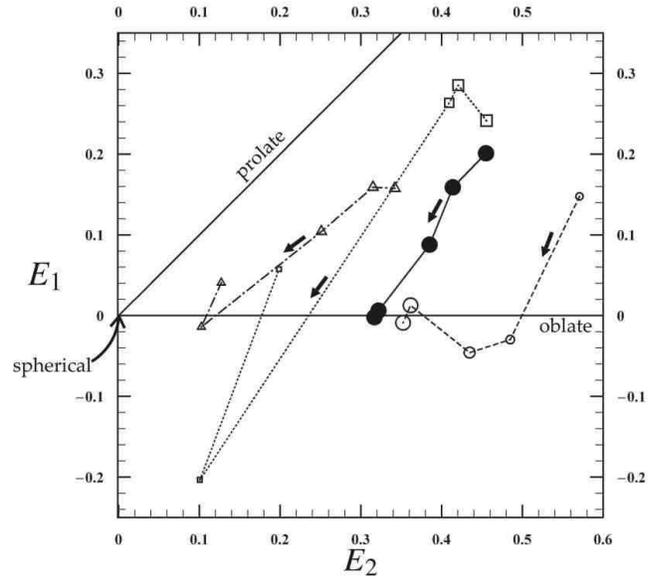}} \caption{The
time evolution of the various populations of the system
(Q,~$m=0.01$) on the plane of the ellipticities
$(E_2,E_1)=(1-\frac{a}{c},1-\frac{b}{c})$, where $a,b,c$ are the
short, intermediate and long axes respectively as calculated by the
moment of inertia tensor of each population. The black dots
correspond to the system (Q,~$m=0.01$) as a whole, the white circles
correspond to the regular component and the rectangles and the
triangles correspond to the chaotic component inside and outside of
the window of the $\log(L_j)$ distribution, respectively. The total
area within each symbol represents the percentage of the associated
population with respect to the total number of particles. The small
arrow in each curve indicates the time arrow. The five snapshots
that are plotted are at $t=(0,60,150,300,600)T_{hmct}$. We observe
that all the different populations exhibit an overall evolution
towards a more oblate configuration ($E_1=0$). However, the regular
component is always far from a spherical distribution (corresponding
to $E_1=E_2=0$), while the latter is more closely approached from
the start by the chaotic orbits outside the window.}
\end{figure}

In order to identify which particles in chaotic motion are most
responsible for the secular evolution of the systems, we collected
the particles with $\log(L_j)$ values in the window
$\log(L_M)-0.3\leq \log(L_j)\leq\log(L_M)+0.3$ in Figs.~3 and 4,
i.e., the particles within roughly one dispersion interval around
the maxima of each distribution. The majority of the particles
inside this window are found to be particles moving mainly in box
orbits in the corresponding smooth centre system before the
insertion of the CM. Up to $t=150T_{hmct}$ these particles
preserve to a large extend their initial, strongly non spherical,
distribution. For example, Fig.~6 shows the distributions of the
particles in regular and in chaotic orbits for the system
(Q,~$m=0.01$) at the snapshot $t=150T_{hmct}$. The left column
shows the projections of the particles in regular orbits on the
three principal planes $x-z$, $y-z$ and $x-y$ as indicated in the
panels. According to Table~\ref{tabt150} these particles
constitute $44\%$ of the total mass of the system. This component
forms a nearly oblate configuration. The middle column shows
similar projections, but for the particles lying inside the $\pm
0.3$ window around the value $\log(L_M)=10^{-1.62}$ of Fig.~3e.
This mass constitutes approximately $36\%$ of the total mass and
has a triaxial strongly elongated configuration. On the other
hand, the right column shows the distribution for the remaining
particles in chaotic orbits (outside the corresponding window).
This mass, which constitutes approximately $20\%$ of the total
mass, creates an almost spherical background.

The source of secular evolution of the system (Q,~$m=0.01$) is the
mass (in chaotic motion) of the middle column. During the
evolution, the majority of the particles' orbits become regular,
mainly SAT orbits (Kalapotharakos et al. 2004; Kalapotharakos \&
Voglis 2005). The particles remaining in chaotic orbits form
finally an almost spherical distribution. This appears in Fig.~7
which shows the particles in regular or chaotic orbits in the same
system but for the snapshot $t=300T_{hmct}$. The particles in
regular orbits constitute approximately $78\%$ of the total mass
(left column). At this time the system has already settled down to
a nearly oblate equilibrium state. It is immediately observed that
the particles in regular orbits (left column) support the form of
an oblate spheroid, while the particles in chaotic orbits (right
column) form an almost spherical distribution.

\begin{figure*}
\label{figq005t150}
\centerline{\includegraphics[width=17cm]{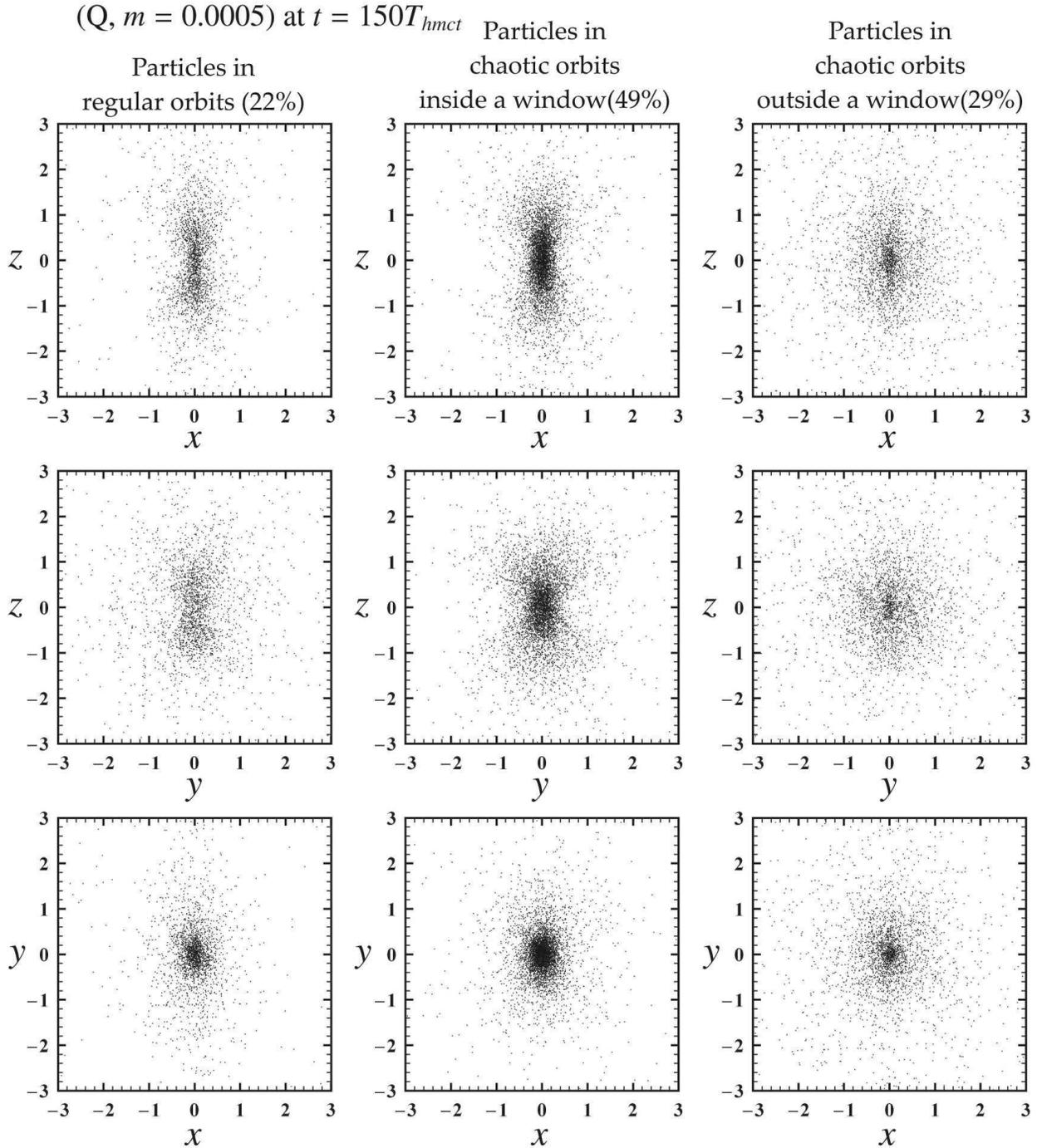}}
\caption{Similar to Fig.~6 but for the (Q,~$m=0.005$) system. The
regular component (left column) constitutes $\simeq 22\%$ of the
total mass and forms a strongly elongated structure. The part of the
chaotic component that lies around the maximum of the distribution
of Fig.~3b (middle column) constitutes $\simeq 49\%$ of the total
mass and has a very similar configuration to that of regular
component. The chaotic diffusion of this mass is mainly responsible
for the secular evolution. The remaining part of the chaotic
component (right column) has a much more uniform and isotropic
configuration. These orbits are already mixed in their available
phase space.}
\end{figure*}

Now, fig.~12 in Kalapotharakos et al. (2004) gives already a hint
that the orbits converted, during the secular evolution, from
chaotic to regular belong to the domain of the distributions of
$\log(L_j)$ near the maximum. Fig.~8 shows a comparison of the
distributions of $\log(L_j)$ for the mass in chaotic orbits of the
system (Q,~$m=0.01$) for three different snapshots, namely $t=0$
(dotted-dashed line), $t=150T_{hmct}$ (solid line) and
$t=300T_{hmct}$ (dotted line). The area below each curve
corresponds to the fraction of the mass in chaotic orbits at each
snapshot. The clear maximum appearing at $\log(L_M)=10^{-1.62}$ of
the first distribution (at $t=0$) is smaller in the second
distribution (at $t=150T_{hmct}$) and even smaller (but still
detectable) in the last distribution (at $t=300T_{hmct}$) at the
position $\log(L_M)\simeq10^{-1.5}$. The decrease of this maximum
implies that many chaotic orbits were converted gradually to
regular orbits. Such conversions drive the secular evolution of
the system, and they are produced mainly from particles in the
area of this maximum.

\begin{figure}
\label{figchamom}
\centerline{\includegraphics[width=6.5cm]{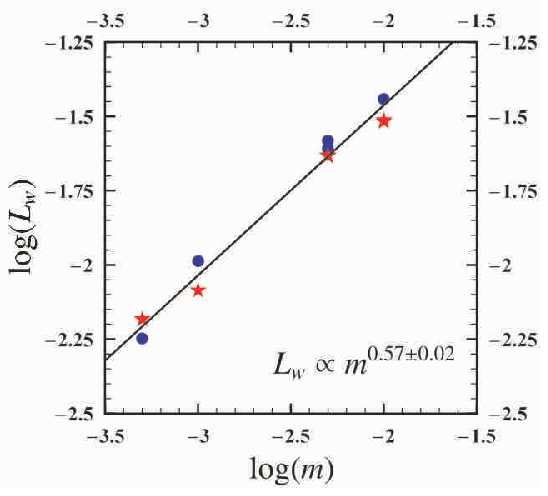}}
\caption{$\log(L_w)$ and $\log(m)$ are related through the power law
$L_w\propto m^{0.57\pm 0.2}$.}
\end{figure}

\begin{figure}
\label{figchamom}
\centerline{\includegraphics[width=8.5cm]{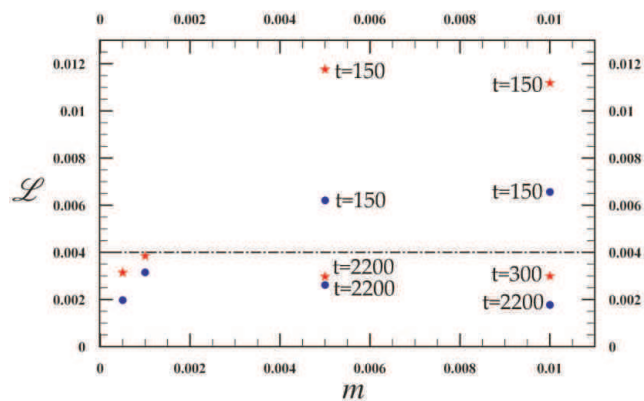}} \caption{The
effective chaotic momentum $\mathscr{L}$ versus $m$. The red stars
correspond to the Q family systems and the blue dots correspond to
the C family systems. In the case of systems showing negligible
secular evolution within a Hubble time one point is plotted for the
value of $\mathscr{L}$ at the time $t=150T_{hmct}$. In the case of
systems showing detectable secular evolution within a Hubble two
points are plotted corresponding to $t=150T_{hmct}$ and to the
equilibrium state. The systems showing negligible or no secular
evolution have $\mathscr{L}$ values lower than $\simeq 0.004$. The
systems with detectable secular evolution ($m=0.005,~m=0.01$) have
initially high $\mathscr{L}$ values ($\mathscr{L}>0.004$) and
finally low $\mathscr{L}$ values ($\mathscr{L}<0.004$).}
\end{figure}

Figure 9 shows how does the distribution of particles in different
types of orbits affect the geometric parameters, i.e. ellipticity
and triaxiality, of the $(Q,m=0.01)$ system, in the course of the
latter's secular evolution. This figure is produced by calculating
the time evolution of the moment-of-inertia tensor of the
$(Q,m=0.01)$ system as a whole, and separately for the regular
orbits, the chaotic orbits within the main window of the
$\log(L_j)$ distribution, and the chaotic orbits outside this
window. The plot shows the ellipticity $E_1=1-b/c$ (vertical axis)
versus $E_2=1-a/c$ (horizontal axis), where a,b,c are the short,
intermediate and long axes respectively as calculated by the
moment of inertia tensor of all the particles of the system (black
circles). The white circles correspond to the same quantities for
regular orbits, the squares to chaotic orbits within the main
window of the $\log(L_j)$ distribution, and the triangles to
chaotic orbits outside the main window. The total area within a
circle, square or triangle symbol is proportional to the
percentage of the associated population with respect to the total
number of particles. The five points per curve correspond to five
different time snapshots as indicated in the figure. Clearly, the
overall evolution of the system is towards an oblate state
(corresponding to the straight line $E_1=0$). All the different
populations exhibit an overall evolution towards a more oblate
configuration. We notice, however, that the regular orbits are
always far from a spherical distribution (corresponding to
$E_1=E_2=0$), while the latter is more closely approached from the
start by the chaotic orbits outside the main window. This further
substantiates the phenomena described so far in Figs.~6 and 7. In
particular, the chaotic diffusion causes the orbits inside the
main window to fill the entire phase space available to them. In
the configuration space, this implies that the chaotic orbits of
any given constant energy fill progressively a larger and larger
volume within the equipotential surface corresponding to that
energy (this is in contrast to the box or tube orbits which only
fill boxes or tubes in such a volume). Furthermore, in typical
galactic potentials the equipotential surfaces are rounder than
the surfaces of equal density. Thus, the chaotic diffusion causes
the spatial distribution of the orbits within the main window to
become more and more spherical and this is the main factor driving
the secular evolution.

In order to demonstrate that the maxima of the distributions of
(Figs.~3, 4) are important as regards the secular evolution of the
systems, Fig.~10 shows a plot similar to Fig.~6, but for the
system (Q,~$m=0.0005$) at the snapshot $t=150T_{hmct}$. The
particles in regular orbits shown in the left column constitute
approximately $22\%$ of the total mass and form a triaxial
strongly elongated configuration, which consists mainly of SAT
orbits and boxlets that have survived after the insertion of the
CM. The particles in chaotic orbits which are inside the window
$(\pm 0.3)$ around the value $\log(L_M)=10^{-2.32}$ (Fig.~3b)
constitute approximately $49\%$ of the total mass (middle column
of Fig.~10). The spatial distribution of this mass is quite
similar to that of the mass of the left column. The right column
shows the projections of the particles in chaotic orbits with
values of $\log(L_j)$ outside the $\pm 0.3$ window (approximately
$29\%$ of the total mass). These particles form a more isotropic
distribution compared to these of the left and middle columns in
the same figure. Because of the low values of $\log(L_M)$, the
rate of chaotic diffusion of the particles in the middle column is
so slow that a macroscopic secular evolution was not detected even
for times much longer than a Hubble time.

In conclusion, the rate of secular evolution of the systems
depends mainly on the fraction of particles in chaotic orbits
which are distributed anisotropically (strongly non spherical) and
on the mean value of the Lyapunov exponents $L_{cuj}$ of the same
orbits, which measures the ability for chaotic diffusion.

The fraction of anisotropically distributed mass in chaotic motion
can be derived by the ratio $\dfrac{\Delta N_w}{N_{total}}$, where
$\Delta N_w$ is the mass inside the window around the maximum of
the $\log(L_{j})$ distributions in Figs.~3 and 4 and $N_{total}$
is the total number particles in the system. If, now, the mean
value of $L_{cuj}$ for the mass inside this window is $L_w$, a
measure of the efficiency of the chaotic diffusion can be obtained
by the quantity
\begin{equation}
\label{eqchamom} \mathscr{L}=\dfrac{\Delta N_w}{N_{total}}L_w
\end{equation}
hereafter called \emph{effective chaotic momentum}. The physical
motivation behind the characterization of $\mathscr{L}$ as a
`momentum' is that the quantity $L_w$ represents a `speed', i.e.
the speed of growth of deviations in the tangent space to the
phase space of the particles' motion, while this speed is
multiplied by a mass $\Delta N_w$, i.e., the total mass of the
particles in chaotic orbits causing substantial chaotic diffusion.

\begin{figure*}
\label{figcoef}
\centerline{\includegraphics[width=17cm]{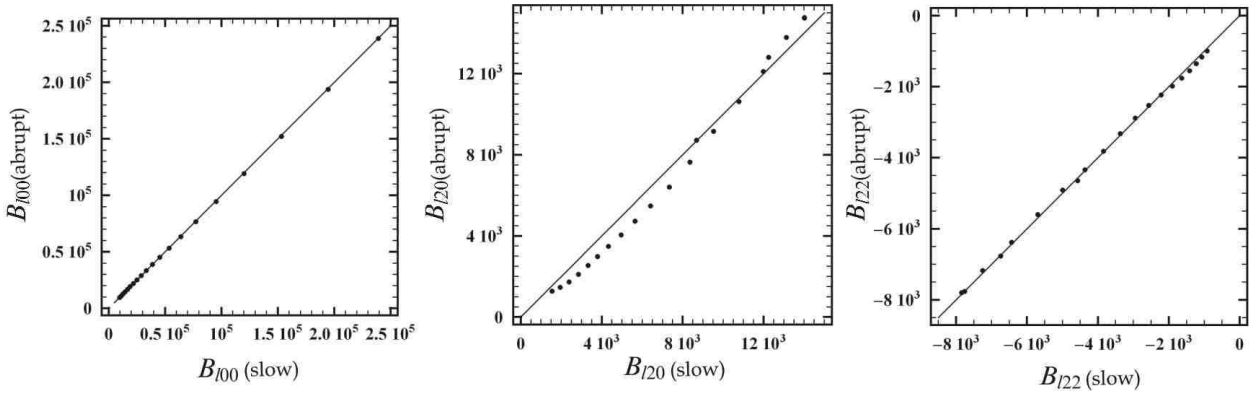}} \caption{A
comparison of values of the most important coefficients
($B_{l00},B_{l20},B_{l22},~l=0...19$) of the multipole expansion of
the potential (see eq.~13 of Kalapotharakos et al. 2004) of the
final equilibrium state of the system (Q~$m=0.01$) when the
introduction of the CM  is done slowly (values in the horizontal
axis) or abruptly (values in the vertical axis). The points in these
diagrams lie close to the diagonal, meaning that the final states
derived by the two simulations are almost the same.}
\end{figure*}

Now, Fig.~11 shows the value of $\log(L_w)$ versus $\log(m)$ for all
the experiments. A power law fit $L_w\propto m^s$ with $s=0.57\pm
0.02$ is found, which is similar to the power laws found in the
previous section. On the other hand, the relation between
$\mathscr{L}$ and the rate of systems' evolution is shown in
Fig.~12. The red stars correspond to the systems of the Q family
while the blue dots correspond to the systems of the C family. For
those systems exhibiting detectable secular evolution within a
Hubble time, i.e. the Q and C systems with $m=0.005$ or $m=0.01$,
the figure shows two points per system corresponding to measurements
of $\mathscr{L}$ at two different snapshots namely $t=150T_{hmct}$
and $t$ equal to time when a system reaches its final equilibrium
state (as indicated in the figure). For the systems with low values
of the CM, Fig.~12 shows only one point corresponding to the value
of $\mathscr{L}$ at the snapshot $t=150T_{hmct}$.

Figure~12 shows that $\mathscr{L}$ is a good measure of the rate of
secular evolution. For example, at $t=150T_{hmct}$ the system
(Q,~m=0.001) has a very high fraction of chaotic orbits ($80\%$),
but the majority of these orbits have low values of Lyapunov
exponents (Figs.~3c,c$'$). On the other hand, at the same time the
system (C,~$m=0.01$) has a much smaller fraction of chaotic orbits
($30\%$) but the majority of them have high values of Lyapunov
exponents (Figs.~4e,e$'$). Thus, the $\mathscr{L}$ value of the
system (C,~$m=0.01$) is higher than the $\mathscr{L}$ value of the
system (Q,~$m=0.001$), In accordance with that, the system
(C,~$m=0.01$) has a much faster rate of secular evolution than the
system (Q,~$m=0.001$). In the case of systems not showing detectable
secular evolution within a Hubble time (Q and C systems with
$m=0.0005$ or $m=0.001$) the value of $\mathscr{L}$ is low. A
threshold value appears to be $0.004$. That is, in all the evolving
systems $\mathscr{L}$ initially takes values higher than the value
$0.004$. As the systems evolve, the value of $\mathscr{L}$
decreases. When the final equilibrium is established, $\mathscr{L}$
falls below the value $0.004$.

\section{Discussion and Conclusions}
This paper deals with a series of $N-$body models representing
elliptical galaxies with central masses. The main addressed
question regards a quantitative characterization of the chaotic
diffusion caused by the insertion of a central mass in systems
containing initially many box orbits, as well as of the
consequences of such a diffusion process in the secular evolution
and macroscopic features of the final states of the systems under
study. The following is a summary of our principal findings:

1) The insertion of the central mass initially converts the
majority of regular box orbits to chaotic orbits. The fraction of
chaotic orbits raises from $20\%-30\%$ to $50\%-80\%$. These `new'
chaotic orbits diffuse in phase space changing the form of a
system, which gradually becomes more spherical (less prolate). Due
to this change, the phase space is gradually transformed as well.
The area corresponding to short axis tube orbits grows while the
chaotic domain is reduced. As the time goes on, many particles in
chaotic orbits are trapped gradually by the tori of SAT orbits and
their orbits are converted to regular. When the secular evolution
ceases the system settles to a new equilibrium state (in most
cases oblate), in which it has a low fraction of chaotic orbits
($\lesssim 25\%$).

One point to be commented here is that, in our experiments, the CM
was "turned on" abruptly while in previous works of the same
subject (e.g. Merritt \& Quinlan 1998, Holley-Bockelmann et al.
2001) the authors introduced a gradual increase of the CM up to
its final value. A question is whether an abrupt insertion affects
significantly the numerical results and in particular the final
state reached by the systems. The systems expected to be more
sensitive on the abrupt insertion are those with the faster rate
of evolution. When the developing time scale of the CM is much
smaller than the system's evolution time scale the difference
between an abrupt and an non-abrupt insertion should be
negligible. In our study we are interested in CMs with developing
times being a small fraction of the Hubble time. Thus, systems
with long evolution times fulfil the above criterion. The more
sensitive system is the (Q,~$m=0.01$) the evolution of which
towards the final equilibrium state takes place in less than a
Hubble time. In order to check the difference by a gradual or abrupt
increase of the CM, we evolved this system through a different
simulation in which the developing time of the CM was taken equal
to $\frac{1}{5}t_{Hubble}$ (following the ansatz suggested by
Merritt \& Quinlan 1998) and compared the results with those of our
original simulation. We checked both the final equilibria and the
times needed for their establishment. One
can easily check that the final states reached by the system with
the CM introduced slowly or abruptly are quite similar, the only
essential difference being that when the CM is introduced slowly
it takes some more delay time for the system to reach the final
state. A persuasive test to see that the final systems are quite
similar is to compare the gravitational potential functions in the
end of the two simulations. This was done by checking the degree of
concentration of the coefficients of the multipole expansion of the potential
(see eq.~13 of Kalapotharakos et al. 2004) of the two systems
(namely the coefficients corresponding to the same radial and
angular quantum numbers) towards the diagonal.
Fig.~13 shows this concentration, which demonstrates that the
final states are quite similar in the two cases, at least macroscopically.

2) We showed that the distributions of the logarithm of the
Lyapunov exponents measured by the inverse radial period of each
orbit show conspicuous maxima consisting of those chaotic
orbits that were previously boxes (before the insertion of the
central mass). These orbits are responsible for the secular
evolution. The Lyapunov exponents of these orbits are related to
the value of the cental mass by a power law relation $LCN\propto
m^s$ with $s$ close to $1/2$. The chaotic orbits around the
maximum are not fully mixed in the phase space, i.e., they have
not yet covered uniformly their available phase space, while for
the chaotic orbits not lying near this maximum evidence is
provided that such a mixing process has already taken place. For
that reason, the chaotic orbits around this maximum have an
elongated spatial distribution, while the remaining chaotic orbits
have a much more isotropic distribution.

An interesting question being raised here regards whether it is
possible to find a physical justification for the power law
scaling $\propto m^{1/2}$ of the Lyapunov exponents of the orbits.
Theoretical work on this direction is in progress and it will be
published in a separate study. We only mention that an answer to
this question can be provided in the framework of the study of
orbits in simplified 3D potentials like
\begin{equation}
\label{toymodel} V(x,y,z)= \underbrace{\frac{1}{2}(\omega_1^2
x^2+\omega_2^2 y^2+\omega_3^2 z^2)}_{V_{harmonic}} +
\underbrace{\epsilon(x z+x y+y^2)}_{quatric~perturbation} -
\underbrace{\frac{m}{r}}_{V_{Kepler}}
\end{equation}
with incommensurable frequencies $\omega_1,\omega_2,\omega_3$. In
such potentials we found that the central mass turns most box
orbits to chaotic, with Lyapunov exponents following precisely the
same scaling, i.e., $\propto m^{1/2}$ as in the simulations with
the galactic potential. Fig.~14 shows an example of this scaling
for orbits with initial conditions taken randomly on a number of
equipotential surfaces of the potential (\ref{toymodel}).

Our preliminary investigation shows that this scaling law can be explained by an
analytical estimation of the eigenvalues of the unstable periodic
orbits formed in the chaotic domain, as a function of $m$
accomplished by the use of perturbation techniques implemented in
the very neighborhood of the unstable periodic orbit solutions.
Namely, the scaling of the Lyapunov exponents with $m$ is related
to the scaling of the eigenvalues of these orbits on $m$.

3) The level of the Lyapunov exponents is an important factor for
one system but it cannot alone determine the rate of the secular
evolution. The rate of secular evolution is significantly
different for each system and depends both on the value of the
inserted central mass and on the original structure of one system.
The value of the central mass, on the one hand, determines the
level of the Lyapunov exponents (through the power law relation),
which are responsible for the diffusion rate. On the other hand,
the system's original configuration determines the fraction of box
orbits that are converted to chaotic orbits able to drive the
secular evolution.

4) A quantity called \emph{effective chaotic momentum} is defined,
that is well correlated with the rate of secular evolution of each
system. The effective chaotic momentum yields the product between
the fraction of particles in chaotic orbits, the chaotic diffusion
of which tends to change the system's form, and the mean Lyapunov
exponent of the same orbits. Neither high values of Lyapunov
exponents nor high fractions of chaotic orbits alone can secure an
effective secular evolution within a Hubble time. An effective
secular evolution needs a proper combination of these two factors.
In particular, when the effective chaotic momentum of a system
falls below the threshold value 0.004 (in the $N-$body units) the
secular evolution practically stops.

\begin{figure}
\label{figosc}
\centerline{\includegraphics[width=8.5cm]{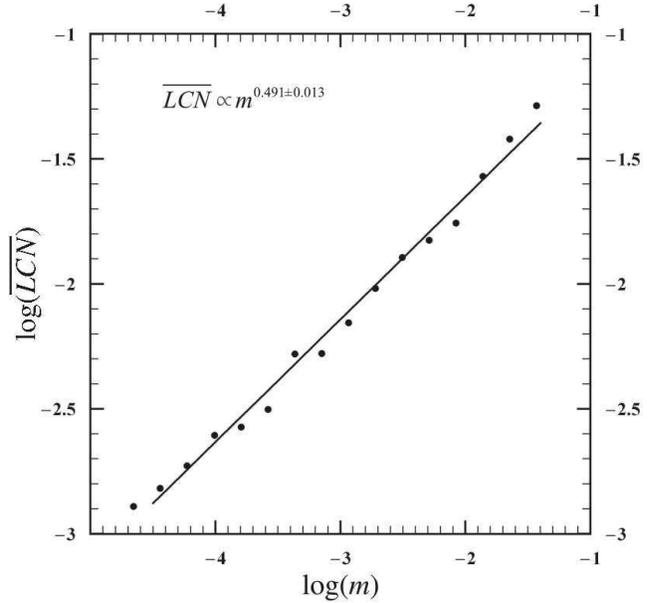}} \caption{The
$LCN\propto m^{1/2}$ relation in the case of the toy model potential
(\ref{toymodel}). For each value of $\log(m)$ we have plotted the
logarithm of the mean value of the Lyapunov exponents
$\log(\overline{LCN})$ of an ensemble of orbits with random initial
conditions in the phase space section corresponding to various
energy levels.}
\end{figure}

\section*{Acknowledgments}

I would like to thank Professor G. Contopoulos for stimulating
discussions and Dr. C. Efthymiopoulos for a careful reading of the
manuscript with many suggestions for improvement. This research
was supported in part by the Research Committee of the Academy of
Athens.

\label{lastpage}

\end{document}